\documentstyle[preprint,aps,eqsecnum]{revtex}
\begin{document}

\draft
\preprint{\vbox{\hbox{U. of Iowa preprint 96-10}}}

\title{
The Oscillatory Behavior of the High-Temperature Expansion
of Dyson's Hierarchical Model:
A Renormalization Group Analysis}

\author{Y. Meurice, S. Niermann, and G. Ordaz }

\address{Department of Physics and Astronomy\\
University of Iowa, Iowa City, Iowa 52246, USA}

\maketitle

\begin{abstract}
We calculate 800 coefficients 
of the high-temperature expansion
of the magnetic susceptibility
of Dyson's hierarchical model with a 
Landau-Ginzburg
measure. 
Log-periodic corrections to the scaling laws appear as in 
the case of a Ising measure.
The period of oscillation appers to be a universal quantity
given in good approximation
by the logarithm of
the largest eigenvalue of the linearized RG transformation,   
in agreement with a possibility suggested by K. Wilson
and developed by Niemeijer and van 
Leeuwen. We estimate $\gamma $ to be 1.300 (with a systematic error
of the order of 0.002)
in good agreement with the results
obtained with other methods such as the $\epsilon $-expansion.
We briefly discuss the relationship between the oscillations
and the 
zeros of the partition function near the critical point in the 
complex temperature plane.
 
\end{abstract}

\newpage
\section{Introduction}

A possible way of testing  our understanding 
of second order phase transitions consists in
calculating the critical exponents as accurately as possible.
Ideally, one would like to use several independent methods and
obtain an agreement within small errors.
The renormalization group method\cite{wilson} 
has provided several approximate methods to calculate the critical
exponents of lattice models in various dimensions.
On the other hand, the same exponents 
can be estimated from the analysis
of high-temperature series\cite{gaunt}.
Showing that these methods give precisely the same answers
has been a challenging 
problem\cite{dis}. In general, one would expect that a well-established
discrepancy could either reveal new aspects of the critical behavior of the 
model considered or point out the inadequacy of some of the methods used.

In order to carry through this program, one needs to
overcome technical difficulties which are specific to the methods
used. An important problem with the high-temperature 
expansion\cite{nickel} is that 
one needs {\it much} longer series than
the ones available\cite{creutz,ising2} 
(which do not go beyond order 25 in most of the 
cases) in order to make precise estimates.
On the other hand, a problem specific to the renormalization group
method is that the
practical implementation of the method
usually requires projections into a manageable
subset of parameters characterizing the interactions. 

It is nevertheless
possible to design a non-trivial lattice model\cite{dyson}, 
referred to hereafter as Dyson's hierarchical model (in order to 
avoid confusion with other models also called ``hierarchical"), 
which can be seen as an 
approximate version of nearest neighbor models and 
for which these two technical difficulties can be overcome.
For Dyson's hierarchical model,
the renormalization group transformation reduces to 
a simple integral equation involving  only the local measure.
This simplicity allows one to control rigorously\cite{sinai} 
the renormalization group
transformation and to obtain accurate 
estimates of the eigenvalues of the linearized renormalization group
transformation\cite{epsi}. More recently, we have shown that
the recursion formula
can be put in a form\cite{high,prl} suitable to the calculation 
of the high-temperature expansion to 
very large order. Consequently,
Dyson's hierarchical model is well suited to compare the
$\epsilon$-expansion and the high-temperature expansion.
Note that unlike the $\epsilon$-expansion, the high-temperature
expansion depends on the choice of a local measure of integration
for the spin variables (e.g. a Ising or Landau-Ginzburg measure).
In order to make this choice explicit when necessary, we will, for 
instance,
speak of Dyson's hierarchical Ising model if we are using a Ising 
measure. 

In a recent publication\cite{prl}, we reported results concerning 
the high-temperature expansion of Dyson's hierarchical Ising model.
We found clear evidence for oscillations in the quantity used
to estimate the critical exponent $\gamma $, called the extrapolated
slope (see section III). When using a log scale for the order in the
high-temperature expansion, these oscillations become regularly 
spaced.
We provided two possible interpretations.
The first
is that the eigenvalues of the linearized renormalization group are
complex. 
The second is that 
the eigenvalues stay real but that 
the constants
appearing in the conventional parametrization of the 
magnetic susceptibility should be replaced 
by functions of $\beta_c -\beta$ invariant
under the rescaling of $\beta_c -\beta$ by $\lambda _1$, 
the largest eigenvalue of 
the linearized renormalization group
transformation. Hereafter, we refer to this explanation as 
``the second possibility".
This second possibility has been mentioned twice
by K. Wilson\cite{wilson} 
and developed systematically by Niemeijer and 
van Leeuwen\cite{nie}.

In Ref.\cite{prl}, we gave several 
arguments against the first possibility. 
A more convincing argument is given in section VII:
explicit calculations of the first fourteen eigenvalues of the linearized
renormalization group transformation not relying on the 
$\epsilon$ or high-temperature expansion show no evidence for 
complex eigenvalues of the linearized transformation. In addition, 
all the results presented below support the 
second possibility.   

In this article, we report the results of calculations of the
high-temperature expansion of the magnetic susceptibility
of Dyson's hierarchical model up to order 800 with a
Landau-Ginzburg
measure. These calculations provide good evidence that the oscillations
appear with a universal frequency given by the second
possibility\cite{wilson,nie} discussed above but with a measure-dependent 
phase and amplitude.
Before going into the technical details related to the 
analysis of the series, we would like to 
state additional conclusions. First, we found
no significant discrepancy between the high-temperature expansion  
and the $\epsilon$
expansion. Second, with the existing methods, 
the high-temperature expansion appears as a rather
inefficient way to estimate the critical exponents of Dyson's hierarchical
model. Third, the 
high-temperature expansion reveals small oscillatory corrections to the
scaling laws which cannot be detected from the study of the 
{\it linearized}
renormalization group transformation. 

These conclusions were reached after a rather lengthy analysis. 
The second possibility introduces potentially an infinite number of
Fourier coefficients and it is useful to first work with simplified
examples in order to develop a strategy to fit the data with
as few unknown parameters as possible. Solvable models
where the second possibility is realized were proposed in 
Ref.\cite{diamond}. 
These models are sometimes called ``Ising hierarchical lattice models"
and should not be confused with Dyson's models.
Further analysis of these models
shows that the zeros of the partition function in
the complex temperature plane are
distributed on the (very decorative) Julia set\cite{mellin}
of a rational transformation. In particular, it is possible 
to relate the oscillations with poles of the Mellin transform located
away from the real axis
at the ferromagnetic critical point. 
In addition, the calculation of the 
amplitude of oscillation for these models 
illustrates a feature which we believe is rather
general, namely that the 
oscillations tend to ``hide'' themselves: 
large frequencies imply (exponentially) small amplitudes. 

This paper is organized as follows. In section II, we specify the models
considered and the methods used for the calculations. In section III,
we explain how to estimate the critical exponent $\gamma$ using
the so-called extrapolated slope\cite{nickel}.
We discuss the effects of the new oscillatory terms on this quantity,
using assumptions which are motivated in subsequent sections.
In section IV, we show that despite a large amplification, the 
systematic and numerical errors 
on the coefficients play no role in our discussion of the oscillations
of the extrapolated slope.  This section also provides a test of our
calculation method in an explicitly solvable case, namely Dyson's
hierarchical Gaussian model.

Inspired by the Ising hierarchical lattice models and the analytical
form of the one-loop Feynman diagrams for Dyson's hierarchical model, 
we designed a simple mathematical function with a singularity
corrected by log-periodic oscillations. This function is defined 
in section V. Its power singularity, as well as the frequency,
amplitudes and phases of oscillations, can be explicitly calculated.
We then show that these quantities can be extracted in good approximation
from a numerical analysis of the extrapolated slope associated with the  
Taylor expansion of the function about a non-singular point.
In section VI, we apply the methods developed in section V to fit the 
extrapolated slope associated with the various high-temperature
expansions calculated. The analysis is complicated by the fact that
the $1/m$ corrections to the large $m$
expansion - $m$ being the order in the high-temperature expansion -
are enhanced by a factor which is approximately 160.
We start with 5 parameter fits, which give robust values for 
the critical exponent $\gamma$ and the frequency
of oscillation $\omega$. From the study of the errors, one can 
design fits with one or two more parameters which have
smaller systematic errors and which are 
reasonably stable under small changes in the fitting interval
or in the initial guesses for the values of the parameters.
  
The results of the numerically stable 
fits are discussed in section VII. The linear relation
between $\omega $ and $\gamma $ predicted by the second possibility
is well obeyed and the value of $\gamma $ is in good agreement with
the value obtained with the $\epsilon-$expansion, which we have checked
using independent methods. 
All results agree within errors of the order 0.002.
We have thus succeeded in finding a theoretical framework in which the new
and existing results fit together. Many questions remain: 
What is the origin of the oscillation? Can we calculate the amplitudes
of oscillation directly? Are similar phenomena present for models
with nearest neighbor interactions? If the example of the solvable
Ising hierarchical lattice models can be used as a guide, these
questions require a better understanding of the susceptibility
in the complex temperature plane.
These questions are briefly discussed in section VIII. In particular, we
give preliminary results concerning 
the zeros of the partition function in the
complex temperature plane which suggests an accumulation of zeros near the 
critical point.

\section{Recursive Calculation of the HT Expansion}

In this section, we describe Dyson's hierarchical model
and the methods used to calculate the high-temperature 
expansion of the magnetic susceptibility.
The models considered here have $2^n$ sites.
Labeling the sites with $n$ 
indices $x_n ..... x_1$, each index
being 0 or 1 , we can write the hamiltonian as
\begin{equation}H=-{1 \over 2} \sum\limits_{l=1}^{n}({c \over 4})^l
\sum\limits_{x_n,...,x_{l+1}}\ (\ \sum\limits_{x_l,....,x_1} \sigma
_{(x_n,....,x_1)} )^2\  \ .
\end{equation}
The free
parameter $c$ which controls the strength
of the interactions is set equal to $2^{1-2/D}$ in order to approximate
a nearest neighbor model in $D$-dimensions. In this article
we {\it only} consider the case $D=3$.
The spins $\sigma_{(x_n,....,x_1)}$ 
are integrated with a local measure which needs to
be specified. In the following we consider 
the 
Ising measure, where the spins 
take only the values $\pm 1$, and measures where the spin variables
are integrated with a weight 
$ e^{-A\sigma ^2-
B \sigma ^4}$, which we call Landau-Ginzburg measures. 
In the particular case $B=0$, we obtain a Gaussian measure.
In the following we have used $A=1/2$ with $B=0.1$ or $B= 1$.

The integrations can be performed iteratively using a 
recursion formula studied in Ref.\cite{sinai} .
Our calculation
uses the Fourier transform of this recursion formula
with a rescaling of the spin variable appropriate
to the study of the high-temperature fixed point\cite{high}. 
It amounts to the repeated use of the recursion formula 
\begin{equation}
R_{l+1}(k)=C_{l+1}\exp(-{1\over 2}\beta 
({c\over 2})^{l+1 }{{\partial ^2} \over 
{\partial k ^2}})(R_{l}({k\over \sqrt{2}}))^2
\  ,
\end{equation}
which is 
expanded to the desired order in $\beta$. 

The initial condition for the 
Ising measure chosen here is
$R_0=cos(k)$. For the Landau-Ginsburg measure, the coefficients in the 
$k-$expansion have been evaluated numerically.
The constant $C_{l+1}$ is adjusted in such a way that $R_{l+1}(0)=1$.
After repeating this procedure $n$ times, we can extract
the finite volume magnetic susceptibility 
$\chi_n (\beta)=1\ + \ b_{1,n}\beta \ + \ b_{2,n} 
\beta ^2 \ + \ ...$ from the Taylor 
expansion of $R_{n}(k)$, which reads $1-(1/2)k^2 \chi _{n}+...$.
This method has been presented for the Ising measure in Ref.\cite{high} 
and checked using results obtained with conventional
graphical methods\cite{jmp}.
In the calculations presented below, we have used $n$=100, which
corresponds to a number of sites larger than $10^{30}$.
The errors associated with the finite volume are negligible 
compared to the errors associated with numerical round-offs
as explained in section IV.

\section{The Extrapolated Slope}

In order to estimate $\gamma $, we will use a quantity called the 
extrapolated slope\cite{nickel} and denoted $\widehat{S}_m$ hereafter.
The justification for this will be made clear after we recall
its definition. 
First, we define $r_m=b_m/b_{m-1}$, the ratio of two
successive coefficients. We then define the normalized slope 
$S_m$ and  
the extrapolated slope $\widehat{S}_m $ as
\begin{eqnarray}
S_m & = & -m(m-1)(r_m - r_{m-1})/(mr_m -(m-1)r_{m-1})\ ;
\nonumber \\ & & \\
\widehat{S}_m & = & mS_m-(m-1)S_{m-1}\ .
\nonumber
\end{eqnarray}

In the conventional description\cite{parisi} of the renormalization group
flow near a fixed point with only one eigenvalue $\lambda_1 > 1$,
the magnetic susceptibility
can be expressed as
\begin{equation}
\chi=(\beta _c -\beta )^{-\gamma } (A_0 + A_1 (\beta _c -\beta)^{
\Delta } +....)\ ,
\end{equation}
with $\Delta=|ln(\lambda _2)|/ln(\lambda _1)$ and  
$\lambda_2$ being the largest of the remaining eigenvalues.
It is usually assumed that these eigenvalues are real.
When this is the case, one finds\cite{nickel} that 
\begin{equation}
\widehat{S}_m =\gamma -1 + B m^{-\Delta} + O(m^{-2}) \ .
\end{equation}
Remarkably, the $1/m$ corrections coming from analytic contributions
have disappeared, justifying the choice of this quantity.
Instead of this monotonic behavior, oscillations with a logarithmically
increasing period were observed in Ref. \cite{prl}.
Eq.(3.3) was then used, allowing
$B$ and $\Delta $ to be complex and selecting the real part of the modified
expression. This 
introduces two new parameters, and 
the parametrization of the extrapolated slope becomes:
\begin{equation}
\widehat{S}_m =\gamma -1 -a_1 m^{-a_2}cos(\omega ln(m) + a_3) \ .
\end{equation}
This parametrization allows one to obtain good quality fits provided
that $m$ is not too small.

This parametrization is compatible with two interpretations. The first one
is that the eigenvalues of the linearized renormalization group are
complex. We have given\cite{prl} several general arguments against 
this possibility and
an explicit calculation reported in section VI makes this possibility
quite implausible.
The second possibility\cite{wilson,nie} we have considered is that 
the eigenvalues stay real but the constants
$A_0$ and $A_1$ in Eq. (3.2) are replaced 
by functions of $\beta_c -\beta$ invariant
under the rescaling of $\beta_c -\beta$ by a factor $(\lambda _1)^l$,
where $l$ is any positive or negative integer. 
This invariance implies that these functions are 
periodic functions 
in $log(\beta _c -\beta)$ with period $log(\lambda _1)$ and 
can be expanded in
integer powers of 
$(\beta _c -\beta)^{i2\pi \over ln(\lambda _1 )}$.
Consequently, we have the Fourier expansion:
\begin{equation}
A_i(\beta _c -\beta)=\sum_{l\in {\bf Z}}a_{il}(\beta _c -\beta)^{i2\pi l \over ln(\lambda _1 )} \ .
\end{equation}

At this point, we have no additional information about
these Fourier coefficients and possible restrictive relations among them.
In the solvable examples\cite{diamond}
where the second possibility is realized,
the Fourier coefficients decrease exponentially with the mode 
number\cite{mellin},
namely $|a_{il}|\propto e^{-u\omega |l|}$ where 
\begin{equation}
$$\omega\ = \ {2\pi \over ln(\lambda _1 )}$$ 
\end{equation}
and $u$ is a positive constant
expected to be of order 1 but usually difficult to calculate. 
If a similar suppression occurs in the problem considered
here, a truncation of the sum over the Fourier mode should
provide acceptable approximations (see section V for an example). 

If we consider the new parametrization of the
susceptibility - with the constants replaced by sums over Fourier modes -
we obtain a parametrization of the HT coefficients 
as a linear combination of terms of the form $(\beta _c -\beta)^z$.
The asymptotic (at large $m$) form of the 
coefficients is obtained from 
\begin{equation}
(\beta _c -\beta )^z=\beta _c ^z \ \sum\limits_{m=0} ^{\infty} {z\choose m} 
(-1)^m({\beta \over \beta _c})^m 
\end{equation}
and the asymptotic form
\begin{eqnarray}
{z\choose m} (-1)^m=&& {m^{-z-1}\over \Gamma (-z)} \nonumber\\ 
\times (1 + {{z + {z^2}}\over {2\,m}} + 
  {{2\,z + 9\,{z^2} + 10\,{z^3} + 3\,{z^4}}\over {24\,{m^2}}} + &&
  {{6\,{z^2} + 17\,{z^3} + 17\,{z^4} + 7\,{z^5} + {z^6}}\over {48\,{m^3}}}+...
) \ .
\end{eqnarray}
From this we obtain the following asymptotic form for the coefficients:
\begin{eqnarray}
b_m=&&m^{\gamma-1}\sum_{l\in {\bf Z}}K_l m^{il\omega }
(1+((\gamma +il\omega)^2-(\gamma +il\omega))/2m +....)\nonumber \\
&&+m^{\gamma -\Delta-1}\sum_{l\in {\bf Z}}L_l m^{il\omega }
(1+((\gamma-\Delta +il\omega)^2-(\gamma -\Delta+il\omega))/2m +....)+... \ ,
\end{eqnarray}
where the $K_l$ and $L_l$ are (unknown) coefficients proportional
to the (unknown) Fourier coefficients.
In the following, we consider the case of truncated Fourier series
where only $K_0$, $K_{\pm 1}$ and $L_0$ are non-zero.
A tedious calculation shows that to first order in $K_1/K_0$, $L_0$ and
$1/m$, and neglecting terms of order $L_0/m$, we obtain
\begin{eqnarray}
\widehat{S}_m=&&\gamma -1 + 2Re\lbrack i(\omega +\omega^3)
m^{i\omega}K_1/K_0 \rbrack \nonumber \\
&&+L_0m^{-\Delta} (\Delta^3 -\Delta) 2Re\lbrack (\Delta-\Delta ^3 -i
\omega +3i\Delta ^2 \omega +3 \Delta \omega^2 -i \omega ^3)
m^{i\omega}K_1/K_0 \rbrack\nonumber \\
&&+m^{-1} Re\lbrack (\omega ^2 +5i\omega ^2 -2i\gamma \omega^3
+7\omega ^4 -2\gamma \omega^4 -i \omega ^5)
m^{i\omega}K_1/K_0 \rbrack \ .
\end{eqnarray}
From the solvable examples, we expect that $|K_2/K_0|$ should be of the same
order as $|K_1/K_0|^2$. The corrections of this order to $\widehat{S}_m$
read
\begin{equation}
$$2Re\lbrack (2i\omega + 8i\omega ^3)K_2/K_0 + 
(4i\omega ^3 -i\omega)(K_1/K_0)^2\rbrack \ .
\end{equation}
These corrections can be important at moderate $\omega $ (see section V).
Importantly, we see that the $1/m$ terms have reappeared. In the case
where $\omega >> 1$, we see that all the oscillating terms are 
approximately in
phase and proportional to 
$Re\lbrack i m^{i\omega}K_1/K_0 \rbrack $.
In the large $\omega $ limit, the $1/m$ corrections are enhanced by
a factor $\omega ^2$ compared to the leading oscillating term. 
This feature will play an important role in the discussion
of section VI.

Before discussing the fits of the numerical values of the extrapolated slope
for the Ising and the Landau-Ginzburg cases,
we will first show that the errors made in the numerical calculations
do not play any significant role and then discuss the fitting strategy
with a solvable example.

\section{The Effect of Volume and Round-off Errors }

In this section, we discuss the errors 
made in the calculation of the coefficients and show that they have no
relevant effect on the
extrapolated slope for the discussion which follows.
There are two sources 
of errors: the numerical round-offs and the finite number of sites.
We claim that with $2^{100}$ sites and $D =3$, 
the finite volume effects
are several order of magnitude smaller than the round-off errors.

From Eq. (2.2), one sees that the leading volume dependence
will decay like $(c/2)^{n}$. This observation can be substantiated 
by using exact results at finite volume\cite{jmp} for low order
coefficients, or by displaying the values of higher order
coefficients
at successive iterations as in Figure 1 of Ref.\cite{high}.
In both cases, we observe that the $(c/2)^{n}$ law works
remarkably well.
For the calculations presented here,
we have used $c=2^{1\over3}$ (i.e. $D=3$) and $n=100$, which gives 
volume effects on the order of $10^{-20}$.

On the other hand, the round-off errors are expected to grow
like the square root of the number of arithmetical operations.
In Ref.\cite{high}, we estimated this number as approximately
$nm^2$ for a calculation up to order $m$ in the high-temperature expansion
with $2^n$ sites. Assuming a typical round-off error in double
precision of the order of $10^{-17}$ and $n=100$, we estimate that 
the error on the $m$-th coefficient will be of order $m\times 10^{-16}$
(or more conservatively, bounded by $m \times 10^{-15}$).
We have verified this approximate law by calculating the coefficients
using a rescaled temperature and undoing this rescaling after
the calculation. 
We chose the rescaling factor to be 0.8482. The rescaled
critical temperature is then approximately 1. This prevents the 
appearance of small numbers in the calculation.
If all the calculations could be performed exactly,
we would obtain the same results as with the original method.
However, for  calculations with finite precision, the two calculations
have independent round-off errors. The difference between the coefficients 
obtained with the two procedures is shown in Fig. 1 and is 
compatible with the approximate law.
From this, we conclude that for $m\leq 1000$, the errors on the coefficients
should not exceed $10^{-12}$.

We are now left with the task of estimating the effects that the errors on
the $b_m$ have on $\widehat{S}_m$. In general,  $\widehat{S}_m$ is a function
of $b_m,\ b_{m-1}, \ b_{m-2},$ and $b_{m-3}$.
The numerical values of the 
derivative of $\widehat{S}_m$ with respect with these four variables
are shown in Fig. 2 for a Ising measure. 
There is clearly a large amplification factor.
From our upper bound on the errors on the coefficients, we conclude
that the errors on the  $\widehat{S}_m$ should be less than $10^{-4}$
and will not play any role in the following.

We have found independent checks of our error estimates.
First, the smoothness of the data for the $\widehat{S}_m$ rules
out numerical fluctuations which would be visible on graphs.
The size of the data for the calculations with a Landau-Ginzburg measure 
allows
a visual resolution of the order between $10^{-3}$ and $10^{-4}$.
Second, we have calculated  $\widehat{S}_m$ in the Gaussian case
where non-zero results are of purely numerical origin.
The results are displayed in Fig. 3 . It shows that the numerical
fluctuations for the Gaussian hierarchical model are smaller than
$10^{-7}$ for $m\leq 200$. 
This small number indicates that our previous estimates are 
conservative. 

The calculation of the large $m$ coefficients requires a lot
of computing time. We found that using a truncation in the expansion
in $k$ at order 100 could cut the computer time 
by a factor of order 100 while having  
very small effects on the values of the
coefficients. If we plot the differences between the 
values obtained with the truncated and the regular method we obtain
a graph very similar to Fig. 3. For $m\leq 400$, the differences 
are less than $4\times 10^{-6}$, which is compatible with 
the numerical errors discussed above. The data for Landau-Ginzburg
 presented here
has been calculated with the truncated method.

\section{Developing Fitting Methods with a Simple Example}

The form of the coefficients given in Eq. (3.5)
involves an infinite number of parameters. 
In order to see how one can obtain reasonable approximations with
a manageable number of unknown parameter, we will first
consider a simple example. One of the simplest examples
of a function with a singularity and a log-periodic behavior
is given by
\begin{equation}
G(x)\ = \ \sum _{n=0} ^{\infty } {B^n \over {1+A^n x}} \ .
\end{equation}
This example has been motivated by the calculations of Refs. \cite{mellin}
and the form of the analytic expressions corresponding to one-loop
Feynman diagrams for the hierarchical model.
For definiteness, we shall only consider the case where $A$ and $B$
are real and $A > B>1$. 

Picking an arbitrary positive value $x_0$ and introducing a new
variable $\beta \ \equiv \ 1-{x\over x_0}$ we obtain the
``high-temperature expansion'':
\begin{equation}
G(x)\ = \ \sum_{n=0}^{\infty} b_m \beta ^m \ ,
\end{equation}
with coefficients
\begin{equation}
b_m\ = \ \sum_{n=0} ^{\infty} {B^n A^{mn} x_0^m \over{(1+A^n x_0)^{m+1}}} \ .
\end{equation}
The critical value of $\beta $ is 1 and is obtained by setting $x=0$ in 
its definition.

Using the Mellin transform technique discussed in Refs. \cite{mellin},
we can rewrite 
\begin{equation}
G(x)\ = \ G_{reg} (x) + G_{sing} (x) \ ,
\end{equation}
with
\begin{equation}
G_{reg}(x)\ = \ \sum _{l=0} ^{\infty} (-1)^l x^l (1-BA^n)^{-1} 
\end{equation}
and
\begin{equation}
$$G_{sing}(x)\ = \ {\omega \over 2} x^{-a} \sum\limits_{p=-\infty} ^{+\infty}
{x^{-ip\omega}\over sin(\pi (a+ip\omega))} $$ \ ,
\end{equation}
where we have used the notation
\begin{equation}
a={lnB\over lnA} 
\end{equation} 
and 
\begin{equation}
\omega={2\pi \over lnA} \ .
\end{equation}
The complex part of the exponents comes from the fact that the 
Mellin transform of $G(x)$ has poles away from the real axis. 
Substituting $(\beta _c -\beta)x_0 $ for $x$ and considering 
$a$ as a critical exponent, the analogy with the 
original problem becomes clear. Neglecting the regular part in (5.4)
and proceeding as in section 3, we obtain the asymptotic
form of the coefficients as in (3.9),
with $\gamma $ replaced by $a$, $L_l=0$ and
\begin{equation}
K_l\ = \ { x_0 ^{-a} \pi \over{\Gamma(a+i\omega l) sin(\pi(a+i\omega l))}} \ .
\end{equation}
For large $|l|$,
the magnitude of the coefficients decreases like $e^{-{\pi\over 2}
 \omega |l|} |l|^{{1\over 2}-a} $. One sees that fast oscillations
have small amplitudes and vice-versa. This makes the oscillations hard to
observe. In order to get an idea of how to obtain suitable truncations
of the expansion given in Eq. (3.5), we have selected the values
$A=3$, $B=10$ and $x_0=1$ and calculated the coefficients with the 
exact formula (5.3). The sums were truncated in such a way that the remainder 
would be less than $10^{-16}$. 
We then started fitting 
the corresponding $\widehat{S}_m$  using Eq (3.1).
We first used a truncation where the Fourier modes
with $|l|\geq 2$ and corrections of order ${1\over m^2}$ were dropped.
We treated $a$, $\omega$ and the complex number 
$K_1\over K_0$ as unknown coefficients
and determined their values by minimizing the sum of the square of the 
errors
with Powell's method. This allowed us to determine the order
of magnitude of $\omega$ and $a$. Plotting the difference between 
the best fit
and the exact values versus the logarithm of $m$ shows oscillations
twice as rapid as the oscillations in the fit. In other words, 
we needed
the $l=\pm 2$ terms. With these terms included and using the data for 
$m\geq 30$, we obtained $\omega =2.727$ and 
$a=0.4772$, in agreement
with the exact values given by Eq (5.7) and (5.8), with three
 significant digits. 
The data and the fit are shown in Fig. 4.
In this simple example, we found that each correction taken into
account improved the quality of the fits.
This is related to the fact that $\omega $ takes a not too large
value. As we now proceed to discuss, 
a substantially larger value of $\omega $ implies a rather 
more complicated situation.
 
\section{Fitting the Extrapolated Slope}

We now discuss the fits of the extrapolated slope for 
Dyson's hierarchical model.
The data is shown in Fig. 5 for the various measures considered.
From the equally spaced oscillations in the
$ln(m)$ variable, one finds immediately that
$\omega$ is approximately 18. 
According to the exponential suppression hypothesis,
this large value makes plausible 
that only the Fourier modes with $|l|\leq 1$ should be kept.
This simplification unfortunately has the counterpart that 
for large $\omega$ the $1/m$ expansion 
is effectively a $\omega^2 /m$ expansion, 
as explained at the end of section III.

To be more specific, the relative strength of the leading oscillations and
 their $1/m$ corrections is approximately ${\omega^2 \over {2m}}$.
 For $\omega=18$, this means that for $m=162$ the leading term and the 
first corrections have the same 
weight. In the example considered in the previous
section, the critical value was $m=4$ and good quality fits in the 
asymptotic region required
considering the data for values of $m$ larger than about ten times
this critical value - which represents dropping only
5 percent of the data. For the hierarchical model, our data is limited
to 5 times the critical value. Consequently, we probably need about
2500 coefficients in order to get results as accurate as in the example
of section V.
Despite this remark, 
an unbiased parametrization of the form
\begin{eqnarray}
\widehat{S}_m =&&\gamma -1 +a_1 m^{-a_2}cos(\omega ln(m) + a_3)
+a_4 m^{-a_2} +
+a_5 cos(\omega ln(m) + a_6)\nonumber\\
&&+a_7 m^{-1}cos(\omega ln(m) + a_8)
\end{eqnarray}
gives very good quality fits provided that we 
disregard the low $m$ data (see below). 
An example of such a fit is displayed in Fig. 6. The difference
between the data and the fit is barely visible for $m\geq 100$.
For $m\leq 100$ - where we do not have any reason to believe in the
validity of the $1/m$ expansion - the frequency is still well fitted
but not the amplitude. 
The assumption that only the Fourier modes with $|l|\leq 1$ should
be retained can be checked explicitly from the fact 
that the differences between the fit
and the data do not show more rapid oscillations (unlike
in the previous section, where the $|l|=2$ modes were important).

If we want to have any chance of using the $1/m$ expansion as a guide,
it is clear that we have to retain the data for $m>m_{min}$, with
$m_{min}$ larger than say 200. Varying $m_{min}$ 
and the initial values of the parameters 
provides a stability test. 
It appears that for 
the 10 parameter fits mentioned above or their restriction to the
8 parameter case where all the phases of the oscillatory terms are
taken equal, 
the values of the fitted parameters depend
sensitively on the value of $m_{min}$ and on the initial values.
In particular, it makes
no sense to check if the independent parameters 
satisfy relations dictated by the analysis of section 3. 

We have nevertheless been able to design a stable procedure with
less parameters. 
To assess the stability, we vary $m_{min}$ between 200 and 400, keeping 
$m_{max}$ at 800. 
The upper value of $m_{min}$ is chosen in such a way that we have at least two
complete oscillations.
We first set $a_4, \ a_5$ and $a_7$
equal to zero, which yields a parametrization of 
$\widehat{S}_m$ as in
Eq. (3.4). These restricted fits do not suffer from
the sensitive dependence mentioned above. We then analyze the
errors as a function of $m$. In all the cases considered, the
difference between the fit and the data is much smaller than the
 amplitude of the oscillations
(for $m\geq 200$)
and can be approximated by a constant plus a negative power of $m$.
Putting together the fit of the extrapolated slope and the 
fit of the differences, we were able to obtain 6 parameter fits
with a good stability and small systematic errors in $\gamma $. 
We now discuss the two cases separately. 

In the Ising case, the decay of the oscillations controlled 
by $m^{-a_2}$ in the
5 parameter fit 
and the decay of 
the errors are both approximately $m^{-0.6}$. We thus decided
to use Eq.(6.1) with
$a_5$ and $a_7$ equal to zero (making $a_6$ and $a_8$
irrelevant). The 6 parameter fits so obtained 
are then reasonably stable
under small changes in $m_{min}$ (see Figs. 8 and 9). 
Nevertheless, a systematic
tendency can be observed:
when $m_{min}$ is varied
between 200 and 400, $a_2$  evolves slowly from 0.67 to 0.57. 
It is conceivable that if we had data at larger $m$, $a_2$
would evolve toward its expected value 0.46.

In the Landau-Ginzburg case,
the value of $a_2$ obtained from the 5 parameter fits is
very small and the amplitude is in first approximation constant.
We thus set $a_1$ and $a_7$ equal to zero while $a_5$
parametrizes the amplitude of the oscillations and 
$a_4$ corrects the systematic errors.
The power $a_2$ does not have the smooth behavior under a change
of $m_{min}$ it had in the Ising case, however it does the job
that it is required to do: the errors are small and do not show
any kind or tilt or period doubling. These errors are displayed on
Fig. 7. Their order of magnitude is $10^{-3}$, which can be used
as a rough estimate of our systematic errors. Statistical
errors due to the round-off errors are visible on the right part
of the graph and clearly smaller by at least one order of magnitude.
We now proceed to discuss the estimation of the most important
quantities ($\gamma $ and $\omega $) from these fits.

\section{Estimation of $\gamma$ and $\omega $ and 
Comparison with Existing Results}

The values of $\gamma $ as a function of $m_{min}$ are displayed
in Fig. 8. The mean values are 1.3023  in the Ising case and 
1.2998 (1.2978) in the Landau-Ginzburg case with $B=1$ ($B=0.1$). 
We conclude that $\gamma = 1.300$ with a systematic error
of the order of $0.002$.
As explained in the previous section, a precise estimation of 
the subleading exponents seems difficult.

Our results can be compared with those obtained 
from the $\epsilon $-expansion\cite{epsi}, namely
$\lambda_1=1.427$ and $\lambda_2=0.85$. These results 
imply $\gamma=1.300 $ and $\Delta=0.46$.
We have checked these results with methods which do not rely on the 
$\epsilon$-expansion or expansions in the renormalized coupling constants.
First, we have adapted 
a numerical method discussed in Refs.\cite{wilson,wilson2}
to the case of the hierarchical model. We obtained $\lambda_1=1.426 $.
Second, we have used a truncated and rescaled\cite{high} version of Eq. (2.2) 
which corresponds to the usual renormalization group transformation. 
Using fixed values of beta and retaining only
terms of order up to $k^{28}$ at each step of the calculation, we were 
able to determine the fixed point and the linearized renormalization group
transformation in this 14 dimensional subspace. Diagonalizing this matrix,
we found  $\lambda_1=1.426 $ and $\lambda_2 =0.853$. The 
corresponding value of $\gamma $ is 1.302.
For both methods, the errors can be estimated by comparing the linearizations
obtained for successive iterations near the fixed point. The order of magnitude
of these errors is 0.001 in both cases. As a by-product, we also found
that all the other eigenvalues were (robustly) 
real, ruling out the possibility
of complex eigenvalues.

We now consider the values of $\omega $.
A distinct signature of the ``second possibility" discussed in Refs.
\cite{wilson,nie,diamond,mellin} is the relation 
\begin{equation}
\omega \ = \ {3\pi \over {ln2}}\gamma \ .
\end{equation}
This relation is well-obeyed by the a-priori independent quantities
used in the fits as shown in Fig. 9.

\section{Open Questions and Conclusions}

We have thus succeeded in finding a theoretical framework in which the new
and existing results appear compatible within errors of the order of 0.002.
In addition, we also have a qualitative understanding of the behavior 
of the extrapolated slope in the low $m$ region.
Many questions remain to be answered. First, we would like
to understand the origin of the oscillations.
If the example of the solvable
Ising hierarchical lattice models can be used as a guide,
the oscillations are due to poles of the Mellin transform located
away from the real axis. These poles are related to
an accumulation of singularities
at the critical point.
We have tried to get an indication
that a similar mechanism would be present for the models considered 
here. As a first step, we have calculated the expansion of
the partition function about $\beta = 1.179$, a good 
estimate\cite{num} of the critical temperature. We have carried
the expansion up to order 10 for $2^n$ sites with $n=6$ to 12.
The zeros are displayed in Fig. 10. It appears that the 
approximate half-circle on which they lay shrinks around the critical
point when the volume increases. It is not clear that the polynomial
expansion is a good approximation. This could
in principle be checked by searching for the exact zeros.
However, this is a much harder calculation because due to the existence
of couplings of different strengths, the partition function cannot
be written as a polynomial in a single variable of the form
$e^{v\beta }$.

The existence of log-periodic corrections to a singular behavior 
seems to be a feature of hierarchically organized systems.
Empirical observations of such a phenomena have been suggested 
as a possible way to predict the occurrence of
earthquakes\cite{saleur} and stock market crashes\cite{freund}.  
Are similar phenomena present for translationally invariant models
with nearest neighbor interactions?
Using the longest series available\cite{ising2} for a nearest neighbor
model, namely the two-dimensional Ising model
on a square lattice we found no clear evidence for regular log-periodic
oscillations comparable to those seen in Fig. 5.
However, the situation is complicated by the existence an antiferromagnetic
point at $\beta =-\beta _c$ . We used an
Euler transformation as discussed in Ref. \cite{nickel}
to eliminate this problem and found no indications of oscillations
having a period that increases with $m$. On the other hand, the zeros of the 
partition function in the complex temperature plane have been studied
\cite{schrock} extensively. The zeros appear on two circles
in the $tanh(\beta )$ plane, one of them going through the ferromagnetic
critical point. Thus it seems incorrect to conclude that any accumulation
of singularities will create oscillations. Approximate 
calculations of the Mellin transform of the susceptibility of the
two-dimensional Ising model could shed some light on this question.

We also would like to be able to calculate the amplitudes
of oscillation with a method independent of the high-temperature
expansion. As explained in the introduction, the study of the linearized
renormalization group transformation does not provide any indications
concerning the oscillations. Up to now the {\it global} properties of
the flows are only accessible through numerical approaches.
The results presented here should be seen as an encouragement to
develop and test global approaches to the renormalization group flows.

We thank V.G.J. Rodgers for his help regarding computational problems.
One of us (Y.M.), stayed at the Aspen Center for Physics during
the last stage of this work and benefited from illuminating
conversations with Martin Block concerning the fits of section VI.

\newpage
\centerline{\bf Figure Captions}
\noindent
Fig. 1: Difference between  the $\widehat{S}_m$ calculated with the 
two procedures explained in the text. The solid line is 
$m\times 10^{-16}$

\noindent
Fig. 2: Derivatives of  
$\widehat{S}_m$ with respect to $b_m$, $b_{m-1}$, $b_{m-2}$ and
$b_{m-3}$ in the Ising case.

\noindent
Fig. 3: 
$\widehat{S}_m$ in the Gaussian case.

\noindent
Fig. 4: 
$\widehat{S}_m$
for the example of section V, with $A=3,\ B=10$ and $x_0=1$
and the fit described in the text.

\noindent
Fig. 5: 
$\widehat{S}_m$
for the Ising model (crosses) and the Landau-Ginzburg model
with $B=1$ (circles) and $B=0.1.$ (squares)

\noindent
Fig. 6:
$\widehat{S}_m$ for the Ising model and a 10 parameter fit.

\noindent
Fig. 7:
Difference between $\widehat{S}_m $ for Landau-Ginzburg
with $B=1$ and  the fit given by Eq.(6.1) with $\gamma = 1.30137$,
 $\omega = 17.716$, $a_1=a_7=0$, $a_5=-0.01084$, $a_6=0.3367$,
$a_4=0.917$ and $a_2$=1.0589.

\noindent
Fig. 8:
$\gamma $ as a function of $m_{min}$, with $m_{min}$
between 200 and 400 by steps of 5  
for the Ising model (circles) and the Landau-Ginzburg model
with $B=1$ (stars) and $B=0.1.$ (squares).

\noindent
Fig. 9:
${{3\pi \gamma}\over{\omega ln(2)}}$
as a function of $m_{min}$ (as in Fig. 8)
for the Ising model (circles) and the Landau-Ginzbug model
with $B=1$ (stars) and $B=0.1.$ (squares).

\noindent
Fig. 10:
The zeros of the partition function in the complex temperature
plane, in the Ising case with from $2^6$ to $2^{12}$  sites. 
The origin on the graph represents the point $\beta =$ 1.179. 
The outer set of point (on a an approximate ellipse) is $n=6$,
the next set $n=7$ etc..

\vfil


\begin{references}
\bibitem{wilson}
K.~Wilson, Phys.\ Rev.\ B. {\bf 4} 3185 (1971) ; 
 Phys.\ Rev.\ D. {\bf 3} 1818 (1971).
%
\bibitem{wilson2}
K. Wilson, Phys.\ Rev.\ D  {\bf 6}, 419 (1972).
%
\bibitem{gaunt}
D. S. Gaunt and
A. J. Guttmann, in {\it Phase Transitions and Critical Phenomena} vol. 3,
C. Domb and M. S. Green, eds., (Academic Press, New York, 1974) ;
A. J. Guttmann, in {\it Phase Transitions and Critical Phenomena} vol. 13,
C. Domb and Lebowitz, eds., (Academic Press, New York, 1989).
%
\bibitem{dis}
There is a large amount of literature on this subject. References can
be found, e.g., in {\it Phase Transitions, Cargese 1980}, M. Levy, J.C. Le
Guillou and J. Zinn-Justin eds., (Plenum Press, New York, 1982). 
\bibitem{nickel}
B. Nickel, Lecture Notes published in Ref. \cite{dis}.
%
\bibitem{ising2}
B. Nickel, Private communication to A. Guttmann, reported on p. 9 in 
third item under Ref. [3].

\bibitem{creutz}
For recent calculations see e.g:
G. Bhanot, 
M. Creutz, U. Glasser and K. Schilling, Phys.\ Rev.\ B  {\bf 49}
, 12909 (1994) . 
%
\bibitem{dyson}
F. Dyson, Comm.\ Math.\ Phys.\ {\bf 12}, 91 (1969) ; 
G. Baker, Phys.\ Rev.\ B{\bf 5}, 2622 (1972). 
%
\bibitem{sinai}P. ~Bleher and Y. ~Sinai, Comm.\ Math.\ Phys.\ {\bf45}, 
247 (1975) ; P.~Collet and
J. P. ~Eckmann, Comm.\ Math.\ Phys.\ {\bf55}, 67 (1977)  and {\it
Lecture Notes in Physics} {\bf 74} (1978) ;
H. Koch and P. Wittwer, Comm.\ Math.\ Phys.\ {\bf106} 495 (1986) , 
{\bf 138}  (1991) 537 , 
{\bf 164} (1994) 627  .
%
\bibitem{epsi}
P. Collet, J.-P. Eckmann, and B.
Hirsbrunner, Phys.\ Lett.\ {\bf 71B}, 385 (1977). 
%
\bibitem{high}
Y. Meurice and G. Ordaz, J.\ Stat.\ Phys.\ {\bf 82}, 343 (1996).
%
\bibitem{prl}
Y. Meurice, G. Ordaz and V.\ G.\ J.\ Rodgers, Phys.\ Rev.\ Lett. {\bf 75}, 
4555 (1995) .
%
\bibitem{nie}
Th. Niemeijer and J. van Leeuwen, in {\it Phase Transitions and 
Critical Phenomena}, vol. 6,
C. Domb and M. S. Green, eds., (Academic Press, New York, 1976). 
%
\bibitem{parisi}
G. Parisi, {\it Statistical Field Theory} (Addison Wesley, New-York, 1988).
\bibitem{diamond}
M. Kaufman and R. Griffiths, Phys.\ Rev.\ B {\bf 24}, 496 (1981) and
{\bf 26}, 5022 (1982).
\bibitem{mellin}
B. Derrida, L. De Seze and C. Itzykson, J.\ Stat.\ Phys.\ {\bf 33}, 559 (1983);
D. Bessis, J. Geronimo, P. Moussa, 
J.\ Stat.\ Phys.\ {\bf 34}, 75 (1984); 
B. Derrida, C. Itzykson and J.M. Luck, Comm.\ Math.\ Phys.\ {\bf 115}, 
132 (1984).
%
\bibitem{jmp}
Y. Meurice, J.\ Math.\ Phys.\ {\bf 36} 1812 (1995).
%
\bibitem{num}
Y. Meurice, G. Ordaz and V.\ G.\ J.\ Rodgers, J.\ Stat.\ Phys.\ {\bf 77},
 607 (1994).
%
\bibitem{saleur}
H. Saleur, C.G. Sammis and D. Sornette, USC preprint 95-02.
\bibitem{freund}
J. Feigenbaum and P. Freund, EFI preprint 95-58.
\bibitem{schrock}
M. Fisher, Lectures in Theoretical Physics vol 12C, Boulder 1965
Univ. of Colorado Press ;
C. Itzykson, R. Pearson and J.B. Zuber, Nucl. Phys. {\bf B220}
415 (1983);  
V. Matveev and R. Schrock, J. Phys. {\bf A 28} 1557 (1995).
\end{references}
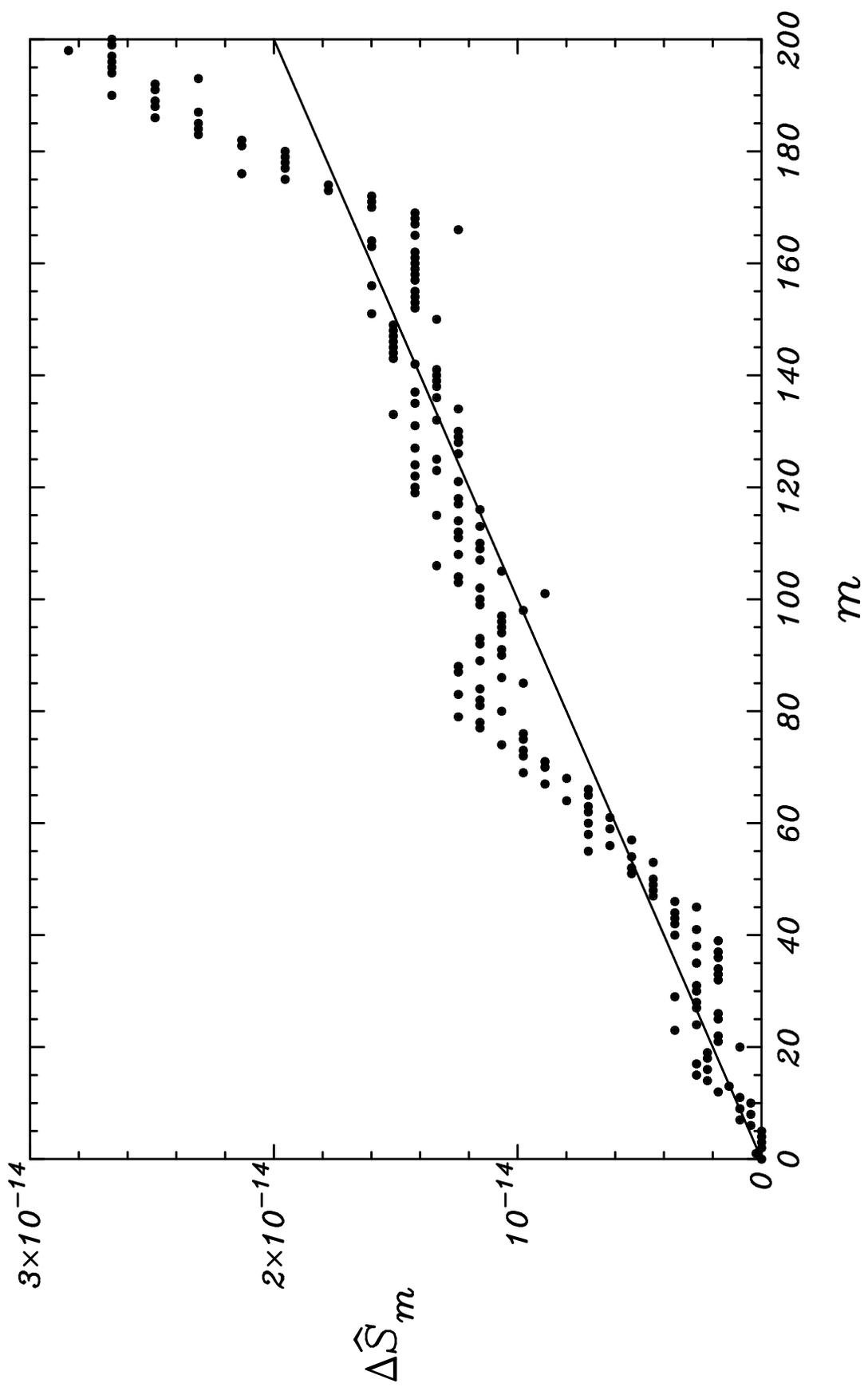
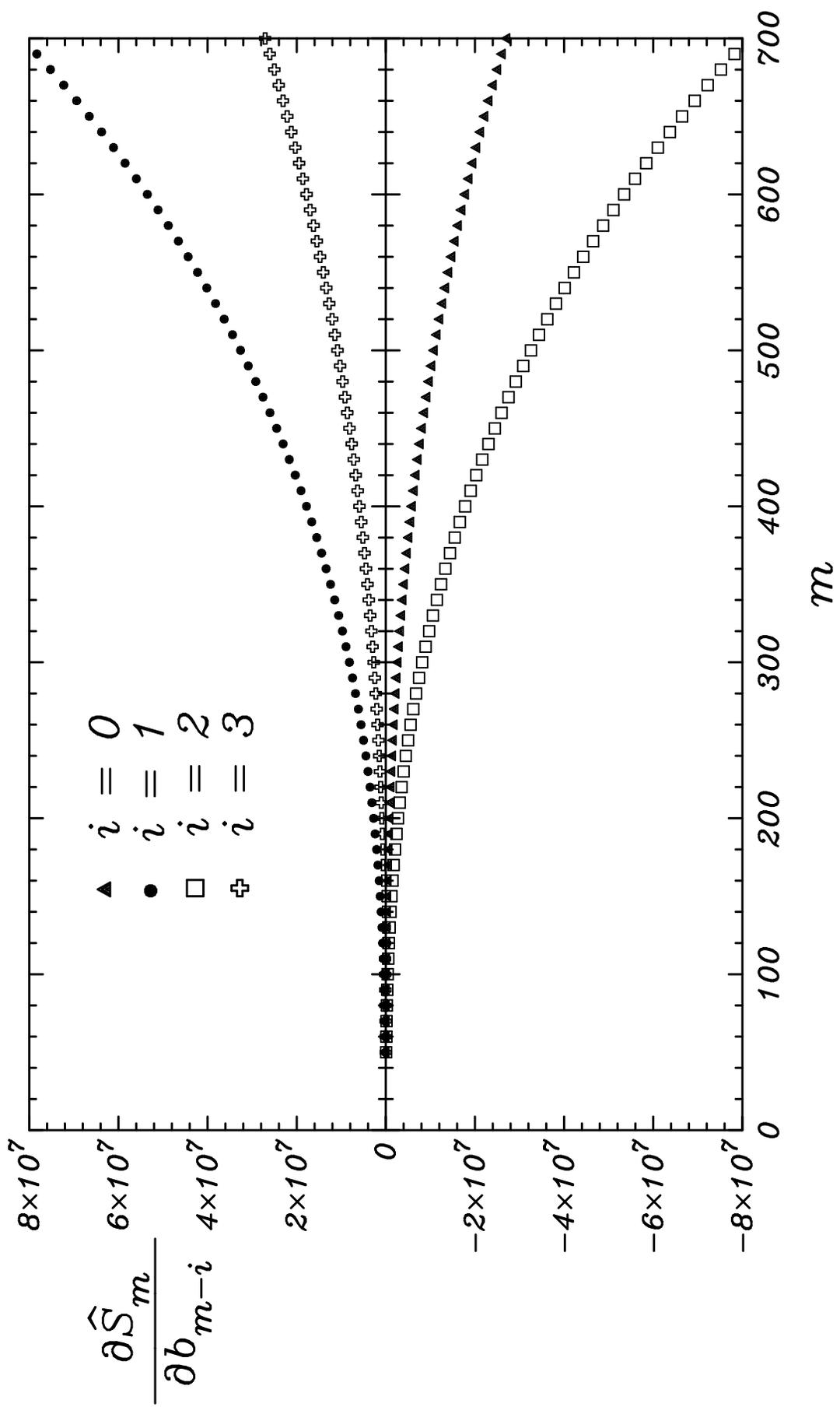
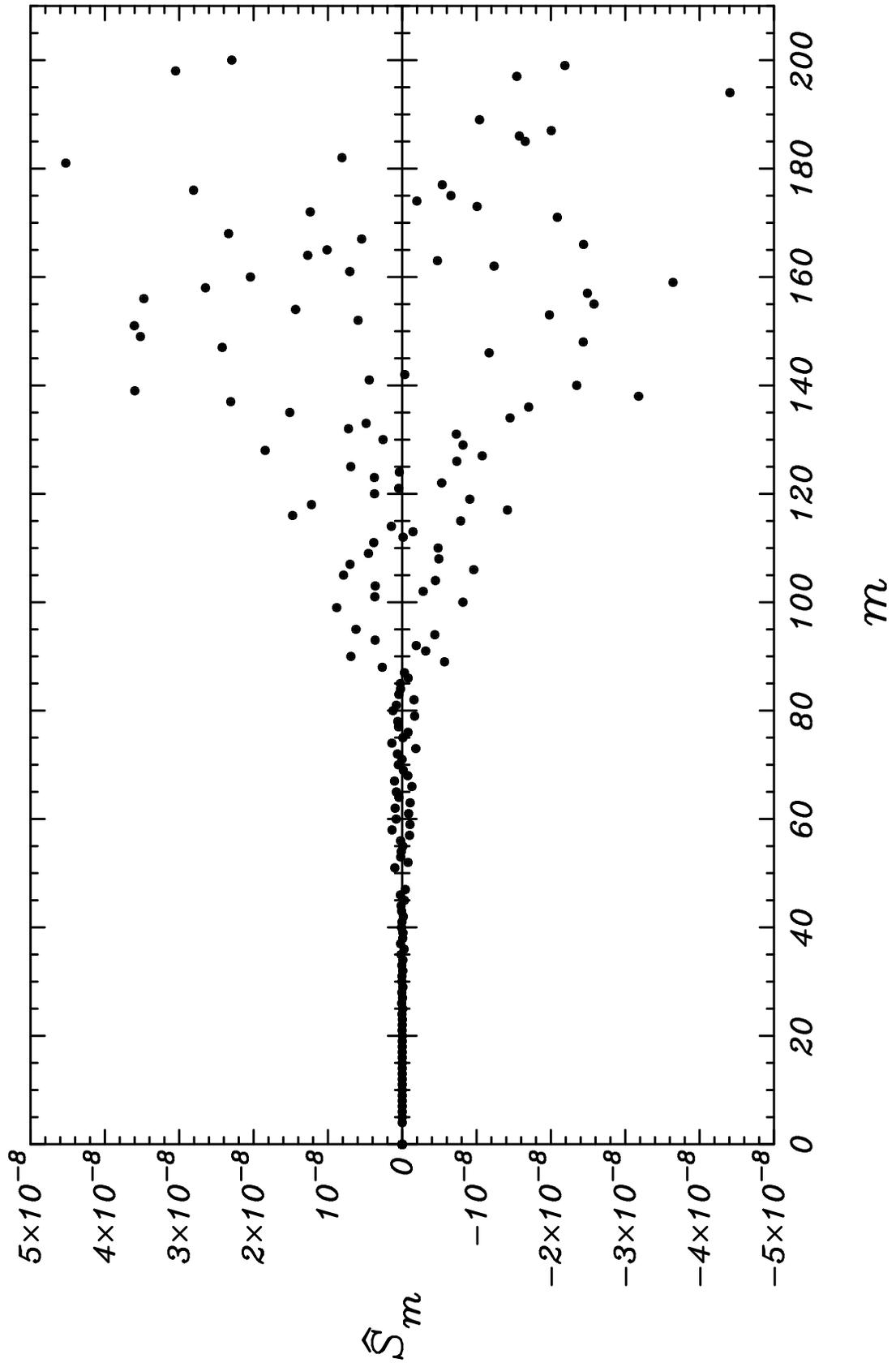
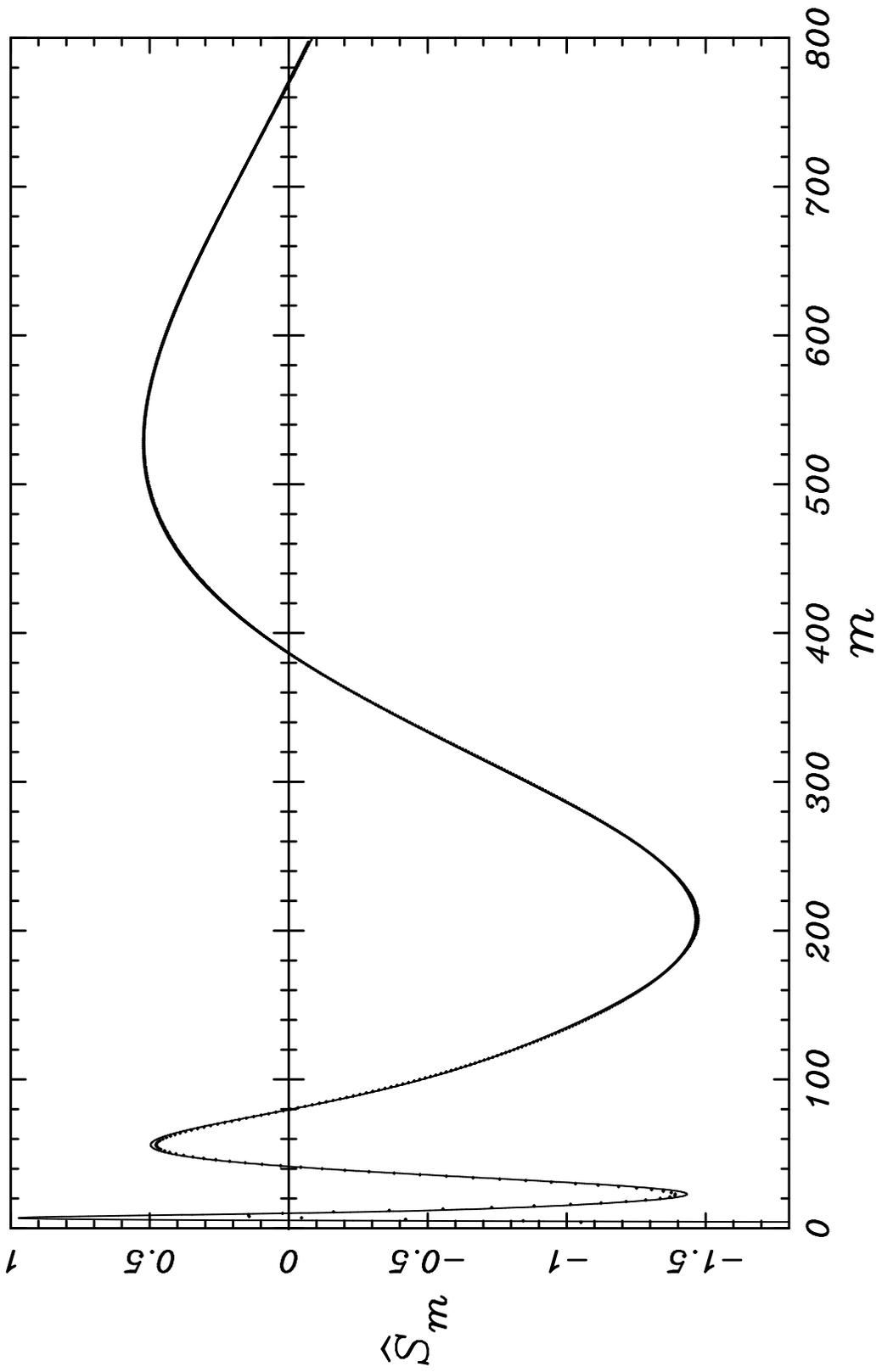
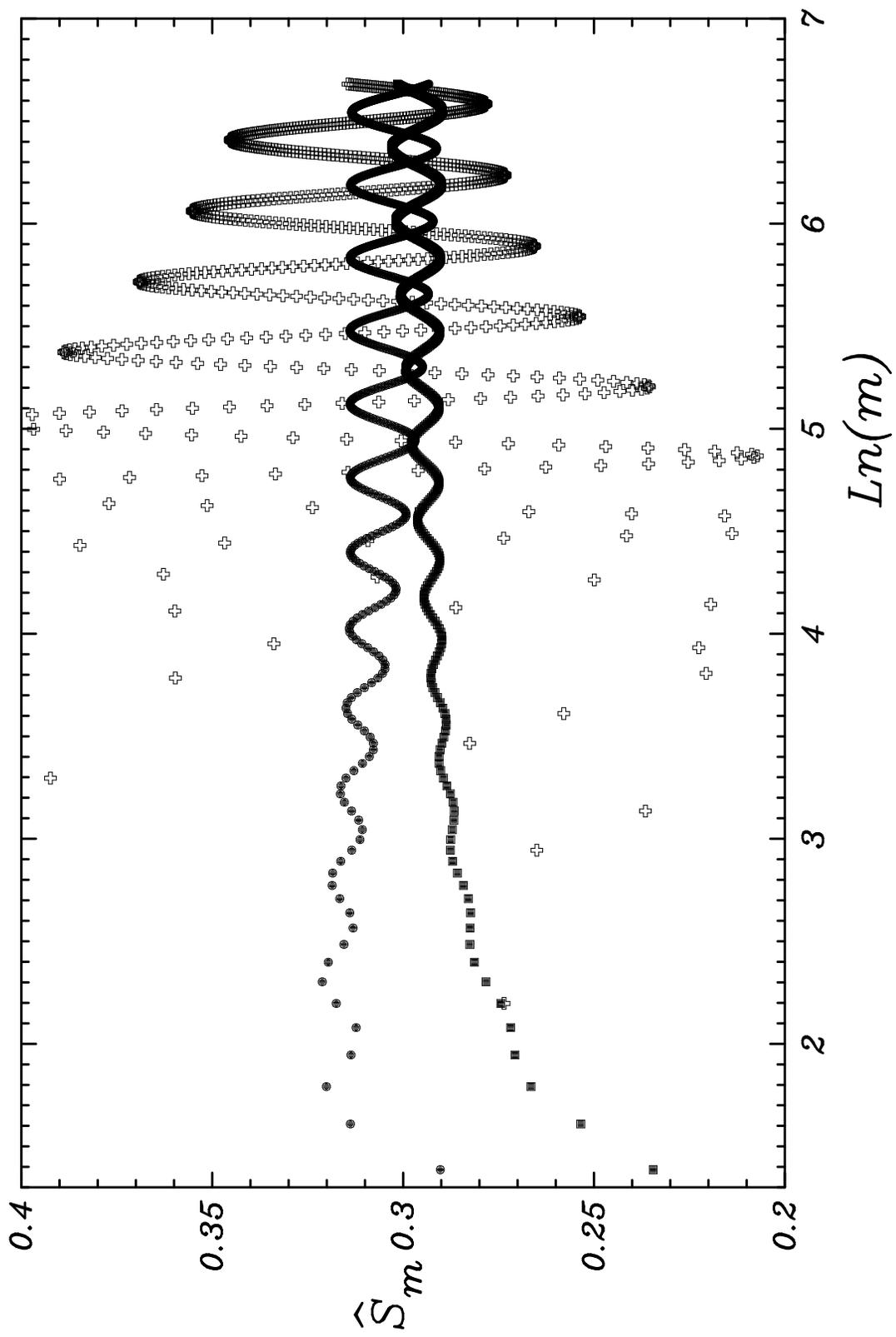
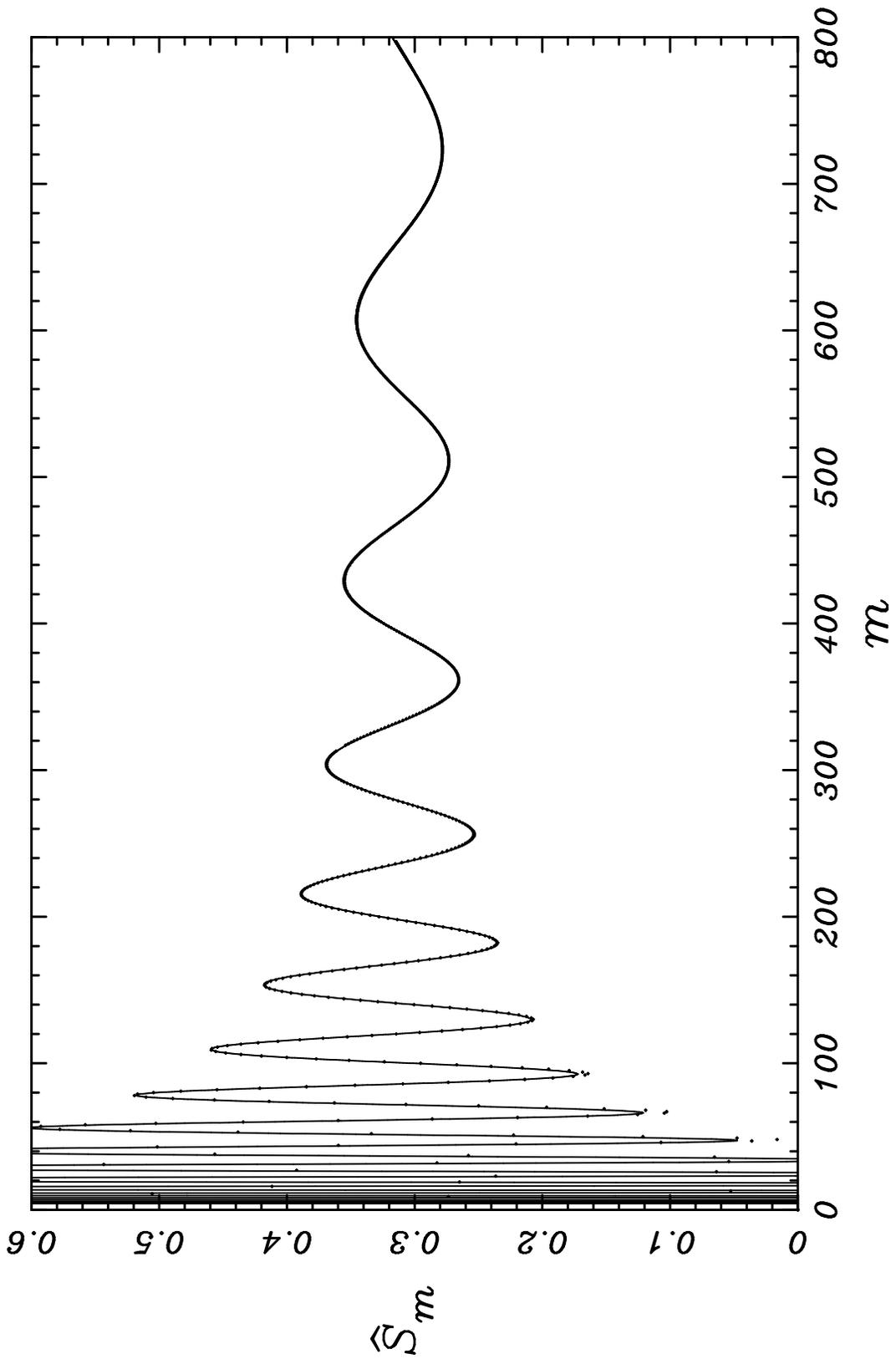
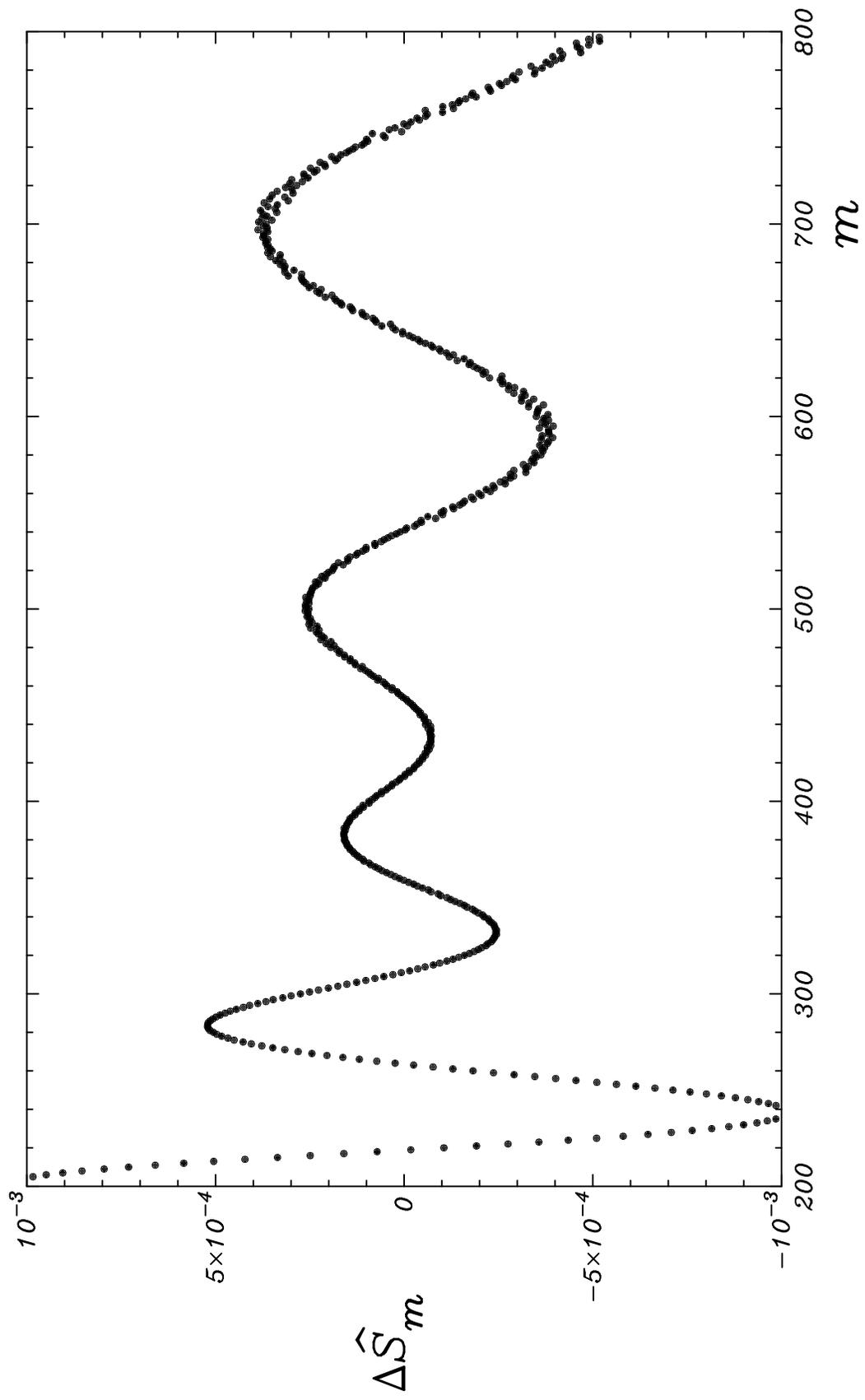
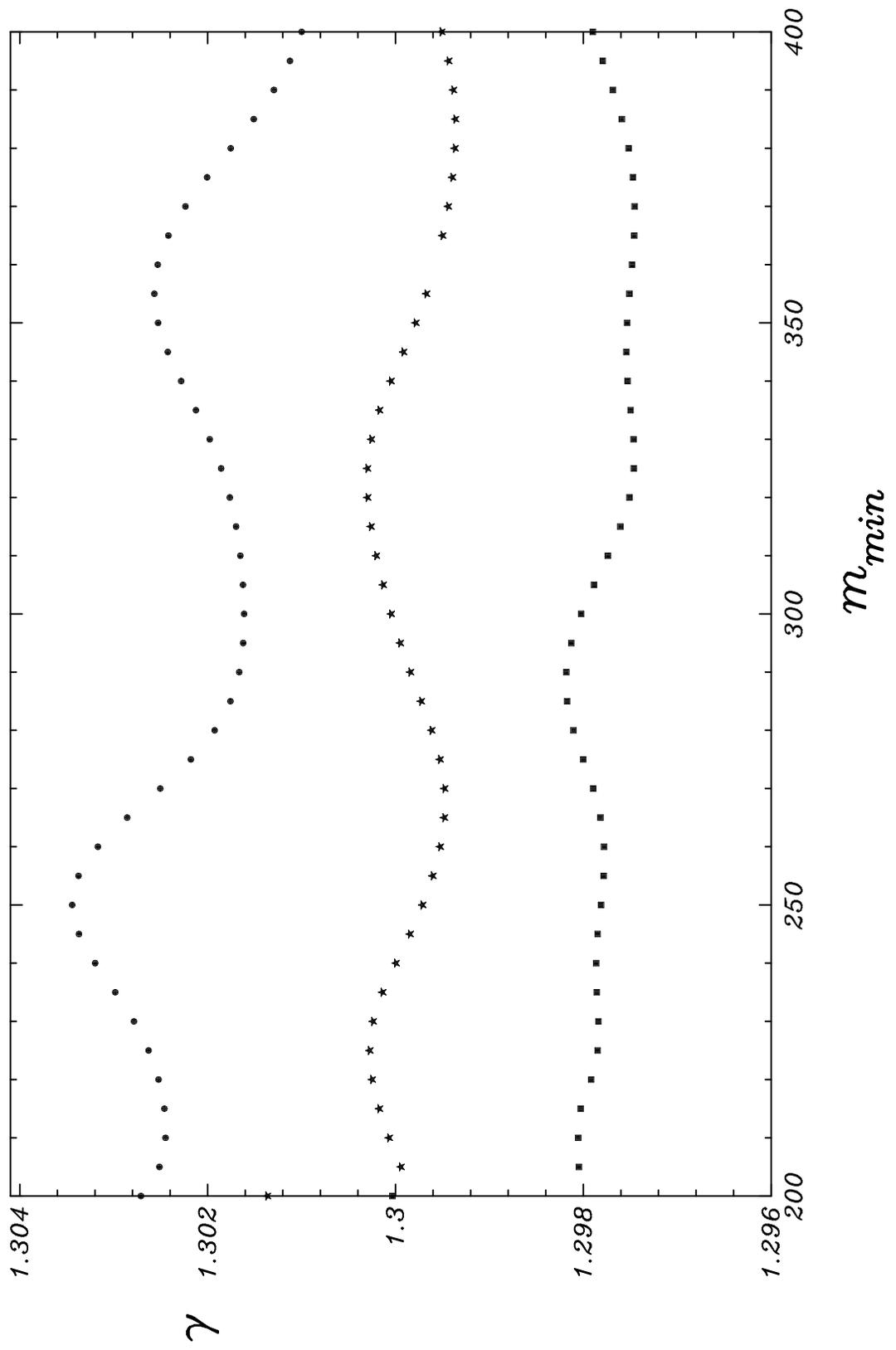
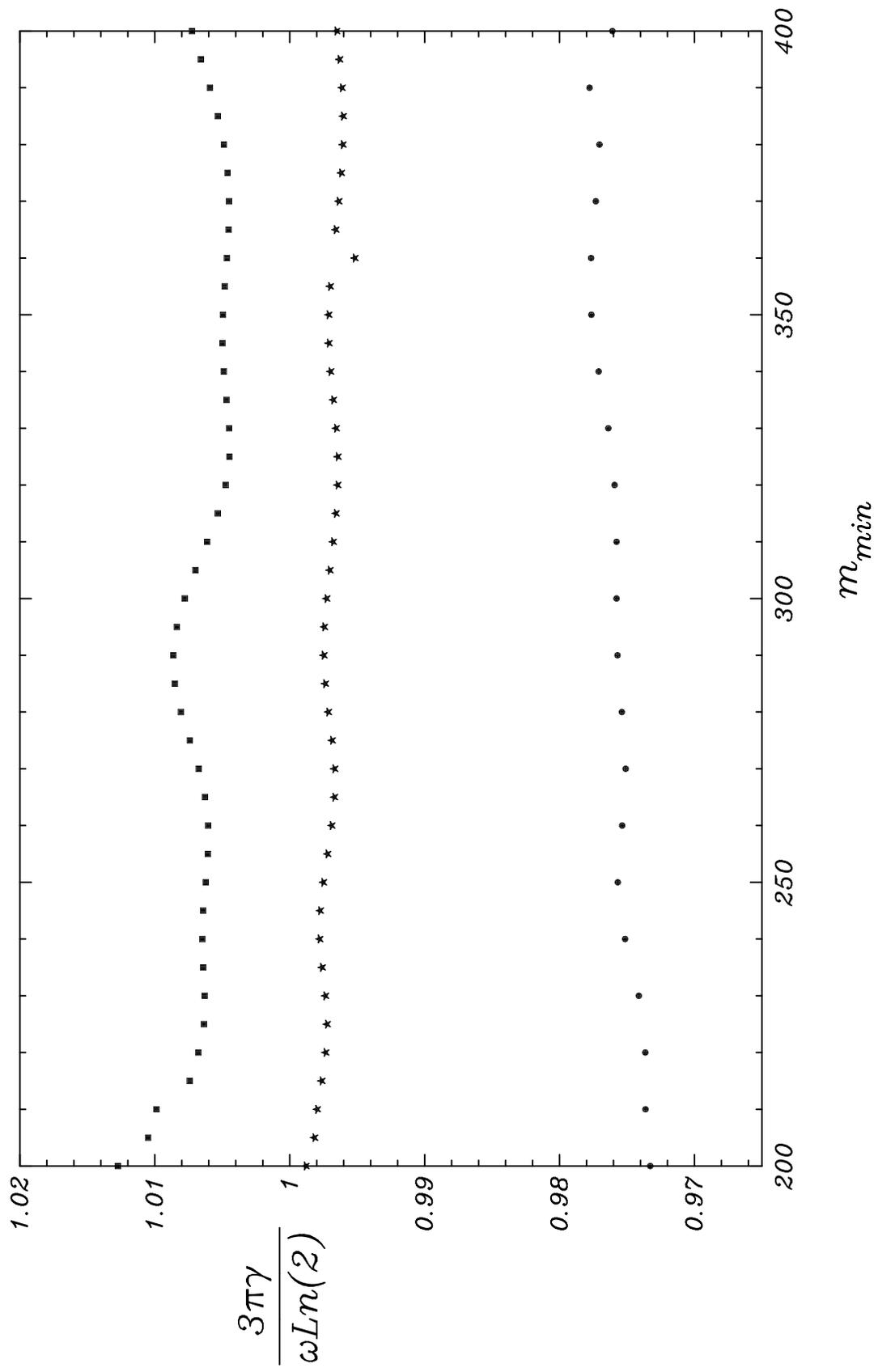
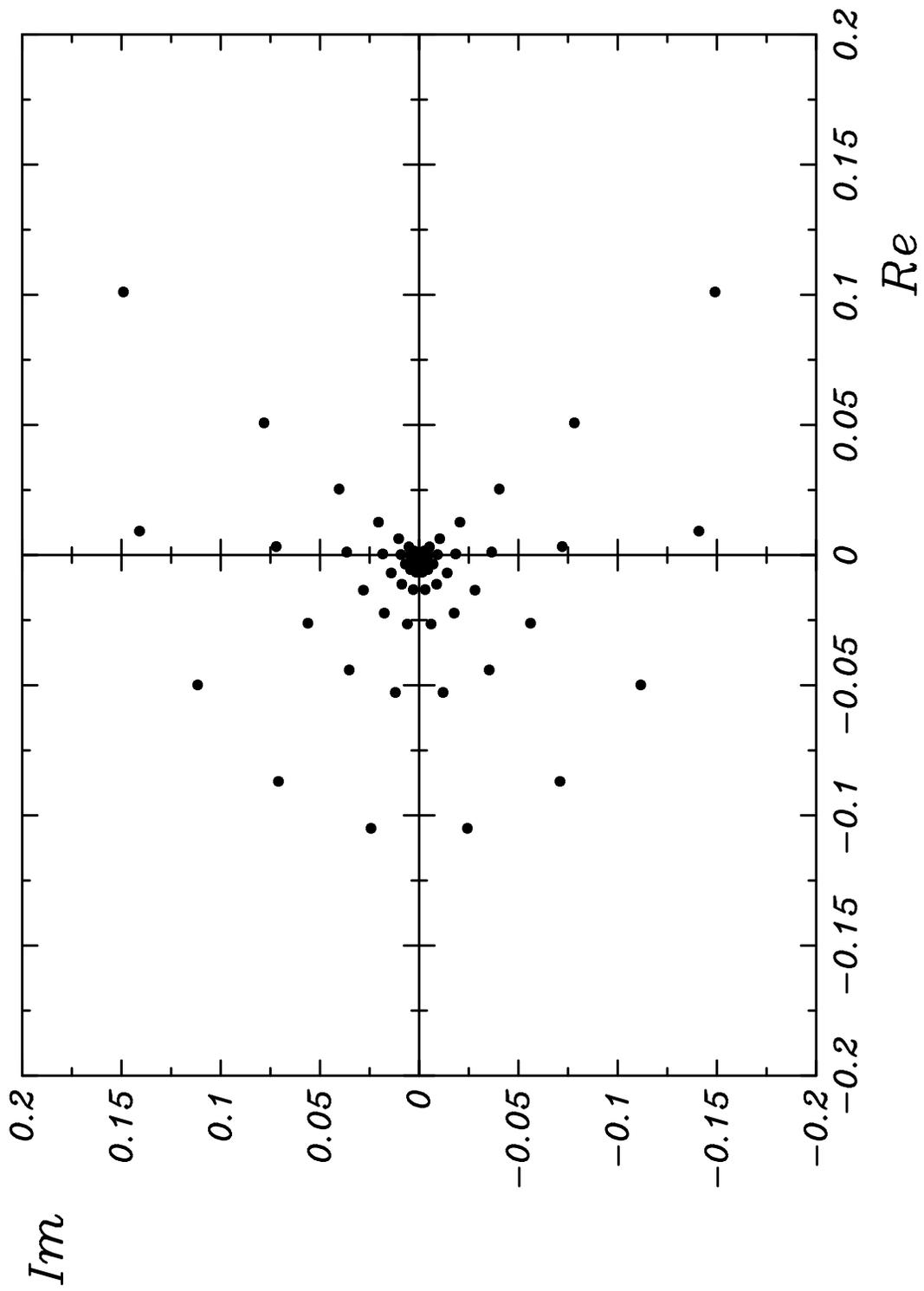
\end{document}